# Green synthesis of silver nanoparticles using *Curcuma longa* flower extract and antibacterial activity

**Kamal Kishor Rajak[1], Pavan Pahilani[1], Harsh Patel[2], Bhavtosh Kikani[2], Rucha Desai[3], Hemant Kumar[1]***

[1]Department of Medical Laboratory Technology, Bapubhai Desaibhai Patel Institute of Paramedical Science (BDIPS), Charotar University of Science and Technology, CHARUSAT Campus, Changa-388421, Gujarat, India.

[2]Department of Biological Sciences, PD Patel Institute of Applied Sciences (PDPIAS), Charotar University of Science and Technology. CHARUSAT Campus, Changa-388421, Gujarat, India.

[3]Department of Physical Sciences, PD Patel Institute of Applied Sciences (PDPIAS), Charotar University of Science and Technology, CHARUSAT Campus, Changa-388421, Gujarat, India.

*Corresponding Author: Hemant Kumar

Department of Medical Laboratory Technology, Bapubhai Desaibhai Patel Institute of Paramedical Science (BDIPS), Charotar University of Science and Technology, CHARUSAT Campus, Changa-388421, Gujarat, India. Phone: +91-9120603548, Email: hemantkumar.cips@charusat.ac.in.

**ABSTRACT:**

Silver Nanoparticles (AgNP's) possess inherent biological potentials that have obliged an alternative, eco-friendly, sustainable approach to "Green Synthesis." In the present study, we synthesized Green Silver Nanoparticles (GAgNP's) using *Curcuma longa L.* (*C. longa*) flower extract as a reducing and capping agent. The synthesized GAgNP's were characterized using UV-Visible spectroscopy, X-ray diffraction (XRD), and High-resolution transmission electron microscopy (HR-TEM), which confirmed their homogeneity and physical characteristics. The GAgNP's were found to contain crystalline silver through XRD and, the particles were confirmed to be homogenous and spherical with a size of approximately 5 nm, as evidenced by UV-Visible spectroscopy, XRD, and HR-TEM. In addition, the biological potential of GAgNP's was evaluated for their antibacterial activities. GAgNP's showed significant activity and form the different sizes of inhibition zone against all selected bacteria *Mycobacterium smegmatis* (*M. smegmatis*) (26 mm), *Mycobacterium phlei* (*M. phlei*) and *Staphylococcus aureus* (*S. aureus*) (22 mm), *Staphylococcus epidermidis* (*S. epidermidis*) and *Klebsiella pneumoniae* (*K. pneumoniae*) (18 mm) and *Escherichia coli* (*E. coli*) (13 mm). The MIC value of GAgNP's was found to be between 625 $\mu$g/mL-39.06 $\mu$g/mL for different microbes tested. With further research, the green synthesis of GAgNP's using *C. longa* flower extracts could lead to the development of effective antibacterial treatments in the medical field.

**Keywords:** Silver nanoparticles, *Curcuma longa,* XRD, HR-TEM, UV-Visible spectroscopy, anti-bacterial

**Graphical abstract**

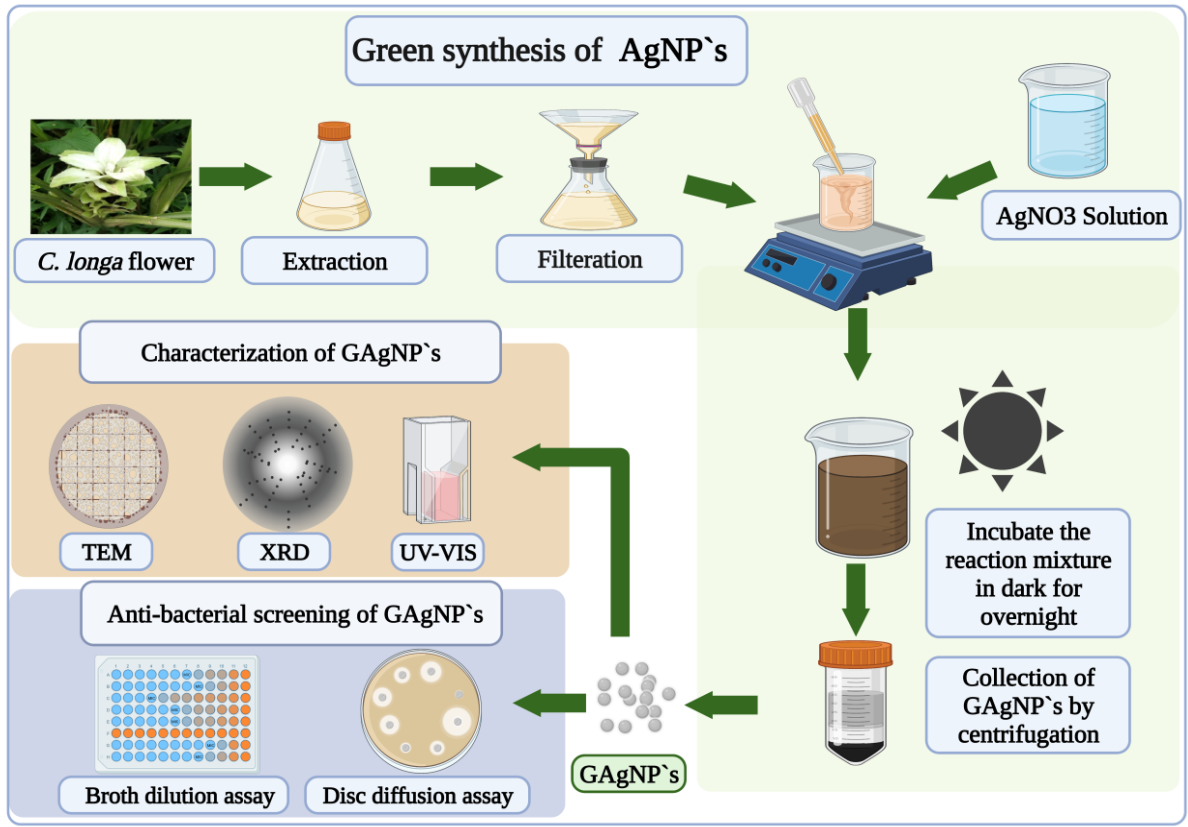

- **INTRODUCTION**

Over the years, nanotechnology has gained a lot of attention due to its ubiquitous applications dedicated to material manipulation on atomic and subatomic levels with dimensions of ≤100 nm. Since 1959, its large-scale use has been recognized by the benevolent community. Manipulated particles are referred to as nanoparticles, which are generally comprised of carbon, metals, metal-oxides, and organic materials.[1] The significance of nanoparticles magnifies due to their larger surface area-to-volume ratio.[2,3]

Nanoparticles are synthesized variably using physical, chemical, and biological approaches. However, some of the mentioned approaches use expensive and toxic chemicals reported to cause harm to ecology and the ecosystem.[4] On the other hand, green synthesis is more sustainable and eco-friendly to counter global environmental problems.[5] Bacteria, fungi, viruses, algae, and plant extracts are the sources of reducing and stabilizing agents in the green synthesis of nanoparticles.[6] Green synthesis of nanoparticles is a promising alternative to chemical and physical methods due to its non-toxic, pollution-free, environmentally-friendly, economical, and sustainable nature.[7] However, there are some challenges associated with the extraction of raw materials, and product quality, as some raw materials, may not be widely available, and the synthesis time may be longer compared to other methods.[8] Despite these challenges, green synthesis remains a promising method for nanoparticle synthesis due to its numerous benefits and potential to address environmental concerns.

Silver nanoparticles hold the crown of most commercialized nano-materials, and their production is anticipated to increase exponentially.[9] Techniques like laser ablation, gamma irradiation, electron irradiation, chemical reduction, microwave processing, and biological synthesis are the reported methods used for their production.[2]

In addition, various research groups reported and described several plant-based biomolecules containing –OH functional group that reduces and stabilize silver ions from $Ag^+$ to $Ag^0$, including proteins, amino acids, enzymes, steroids, alkaloids, polyphenols, quinones, tannins, saponins, carbohydrates, flavonoids, and vitamins.[10–12] With an interdisciplinary approach and an amalgamation of electronics in biology and medical sciences, the exploitation of silver nanoparticles found applications in diagnosis, therapeutics, pharmaceuticals, and biosensors.[13,14] Several recent reports describe the efficacy of green synthesized silver nanoparticles (GAgNP's) against pathogens along with anti-inflammatory and anti-cancer activity against a wide range of cancer cell lines.[15,16]

*Curcuma longa L.* is commonly known as turmeric, Indian Saffron, or the Golden Spice of India. It is a tropical, perennial, monocotyledonous, herbaceous plant that reaches up to 1 meter in height and bears cylindrical, aromatic, and yellow to orange-coloured rhizomes. The plants also have the characteristics of white to green and tinged reddish-purple stem bracts with tapered upper ends at the top of the inflorescence.[17] It is a common spice, a colouring source, and a food preservative.[18] It is a native plant of South Asia and is extensively cultivated in all parts of India. The plant has defined and documented for various ethnobotanical applications.[19] The applications include usage in hepatomegaly and other hepatic conditions, splenomegaly, stomach ulcers, diabetes, cough, chest pain, skin diseases, boils, blood ailments, allergy, and for rheumatism.[18,19] Building upon this, in the present study, we synthesized GAgNP's using the flower of *C. longa.* through the fundamentals of green synthesis and physically characterized the nanoparticles. Further, the GAgNP's were subjected to antimicrobial activity assessment with various bacteria to evaluate their biological efficacy.

- **METHODOLOGY**

**Preparation of Plant Extract.**
The flower of *C. longa* was collected from the village area of Rajnandgaon district (21° 6′ 0″ N, 81° 1′ 48″ E) of Chhattisgarh state in India. The plant material was washed under running water and chopped into small pieces followed by drying in shaded conditions. The dried flower was crushed into a homogenized fine powder, and 10 gm (w/v) of the powdered sample was drenched in 100 mL distilled water and kept under shaking conditions overnight. The immersed solution was filtered using Whatman filter paper no.1, and the extract was collected and stored at 4 °C until further use.

**Synthesis of GAgNP's.**
Bioreduction of Ag ions was carried out using variable $AgNO_3$ concentrations ranging from 1 mM to 5 mM while keeping a constant volume of *C. longa* plant extract for the synthesis of nanoparticles. The concoction was optimized by incubation under dark conditions overnight at room temperature, and the change in colour from pale yellow to reddish brown indicated the reduction of $Ag^+$ to $Ag^0$. The particles were then extracted by centrifugation and were washed thrice with deionized water to remove $Ag^+$ ions.

**Physical characterization of GAgNP's.**

The physical characterization of GAgNP′s in this study involved three techniques: UV-Visible Spectrophotometry, X-Ray Diffraction (XRD) analysis, and High- resolution transmission electron microscopy (HR-TEM) analysis.

*UV-Visible Spectrophotometry Analysis.*

Green synthesized silver nanoparticles were subjected to UV-Visible Spectrophotometry with wavelengths ranging from 200 nm to 800 nm, with distilled water as a reference.[20] This technique provides information on the size and shape of the nanoparticles, as well as their stability and concentration.

*X-Ray Diffraction (XRD) Analysis.*

The crystal structure and size of nanoparticles of GAgNP′s were analyzed by Powder X-Ray Diffractometer (Model: D2- phaser, Bruker AXS, Germany). The measurement was performed over a range of 20° to 90° with a step size of 0.02°/min at a voltage of 30 kV and 10 mA. XRD analysis can provide information on the size, shape, and crystal structure of nanoparticles.

*High-resolution transmission electron microscopy (HR-TEM) Analysis.*

HR-TEM analysis was used to visualize the size, shape, and morphology of GAgNP′s using the FEI Technai F20 TEM instrument. To prepare a thin coat of the sample, the GAgNP′s solution was placed over the carbon-coated copper grid for about one minute, then the solvent evaporated under vacuum, before being sequentially arranged in a grid box subjected to observation. HR-TEM can provide information on the size, shape, and morphology of nanoparticles at a high resolution.

**Antimicrobial activity.**

*Tested Microorganisms.*

The effect of GAgNP′s on pathogenic Gram-positive and Gram-negative bacteria and non-pathogenic acid-fast bacteria were investigated. The strains used in this study were *S. epidermidis* and *S. aureus* (clinically isolated pathogenic Gram-positive strains), *K. pneumoniae*, and *E. coli* (clinically isolated pathogenic Gram-negative strains) and *M. smegmatis* mc$^2$155 and *M. phlei* MTCC 1724 (non-pathogenic acid-fast strains).

*Growth Conditions.*

All Gram-positive and Gram-negative bacteria were cultivated in Muller Hinton broth (MHB) and agar (MHA) at 37 °C for 16 h. Whereas, the acid-fast bacteria were cultured using Middlebrook 7H9 agar supplemented with 10% Oleic acid, albumin, dextrose, and catalase (OADC) and Middlebrook 7H9 broth supplemented with 10% albumin, dextrose, and catalase (ADC) at 37 °C for 48 h (for *M. smegmatis*) and 120 h (for *M. phlei*).

*Disc Diffusion Assay.*

Primary antimicrobial screening of GAgNP's and *C. longa* flower crude extracts was performed by the disc diffusion assay, as described by Patel et al., with slight modifications.[21] Briefly, each strain was cultivated on its respective agar plate, and a single colony of the grown bacteria was used to prepare 0.5 McFarland standard ($1.5 \times 10^8$ CFU mL$^{-1}$) in 1% peptone. Sterile cotton swabs were used to seed the bacterial suspensions on the agar plates. The paper discs (6 mm in diameter) were individually loaded with the 5 mM AgNO$_3$ (20 $\mu$L), GAgNP's (2 mg) and *C. longa* crude extract (2 mg). Streptomycin (STM) (10 $\mu$g) was used as a positive control for Gram-positive and Gram-negative bacteria and isoniazid (INH) (10 $\mu$g) was used for acid-fast bacteria as a positive control. 100% dimethyl sulfoxide (DMSO) (20 $\mu$L) was loaded separately in the control disc as solvent control. The plates were incubated at 37 °C for 16 h (for Gram-positive and Gram-negative bacteria), and at 37 °C for 48 h (for *M. smegmatis*) and 72 h (for *M. phlei*). After incubation, the diameter of the inhibition zone was measured with vernier caliper and reported in millimeters (mm).

*Minimum Inhibitory Concentration (MIC) Determination.*

The MIC was determined by a micro-broth dilution assay as described by Wiegand et al.,[22] The micro-broth dilution assay required $10^6$ cells per mL, which is to be prepared by diluting $1.5 \times 10^8$ cells in growth media. All 96 wells of 96-well microtiter plates were filled with 100 $\mu$L growth media, followed by serial dilutions of GAgNP's and plant crude extract (from 5000 $\mu$g/mL to 39.06 $\mu$g/mL). A 100 $\mu$L solution of diluted bacterial suspension was then added to each well. STM was used as positive control for Gram-positive and Gram-negative bacterial strains, whereas, INH was used as a positive control for acid-fast strains. DMSO was used as a solvent control. The plates were incubated at 37 °C for 16 h for the Gram-positive and Gram-negative bacteria, 48 h for the *M. smegmatis* bacteria, and 72 h for the *M. phlei* bacteria. After incubation, 20 $\mu$L of 0.02% w/v resazurin solution was added to each well, and the plates were incubated until the resazurin colour changed from blue to pink. All the experiments were done in triplicates to ensure the replicability of the experiment.

- **RESULTS & DISCUSSION**

Despite the well-established characterization of the root and leaf of *C. longa,* the pharmacognostic fundamentals of the flower are yet to be thoroughly defined. Ethnobotanical and traditional practices of herbalists intrinsic to the Rajnandagaon district of Chhattisgarh Province in Central India (21.0976° N, 81.0337° E) involve using *C. longa* flowers for curing jaundice. Although the phytochemical composition of *C. longa* flower extract showed a slightly different chemical composition compared to its tuber part (unpublished data), our primary antimicrobial screening using both ethanolic and methanolic extracts on *M. smegmatis,* did not show any inhibitory effects (unpublished data). As a result, we employed the water extract of the *C. longa* flower as a reducing agent to synthesize silver nanoparticles.

**Green Synthesis of Nanoparticles.**
In the present study, GAgNP's were synthesized from the aqueous floral extract of *C. longa*, as shown in **Figure 1 (a)**, which acted as a reducing agent. In order to synthesize the nanoparticles, five different concentrations of silver nitrate were used with the constant volume of plant extract, and the mixture was incubated overnight at room temperature. The primary identification of silver nanoparticles was confirmed by the change in colour of the solution from pale yellow to reddish-brown colour, whereas the control $AgNO_3$ solution (without the flower extract) showed no colour change, as seen in **Figure 1(b)** and **Figure 1(c)**. this colour change may have occurred due to surface plasmon vibration.[23]

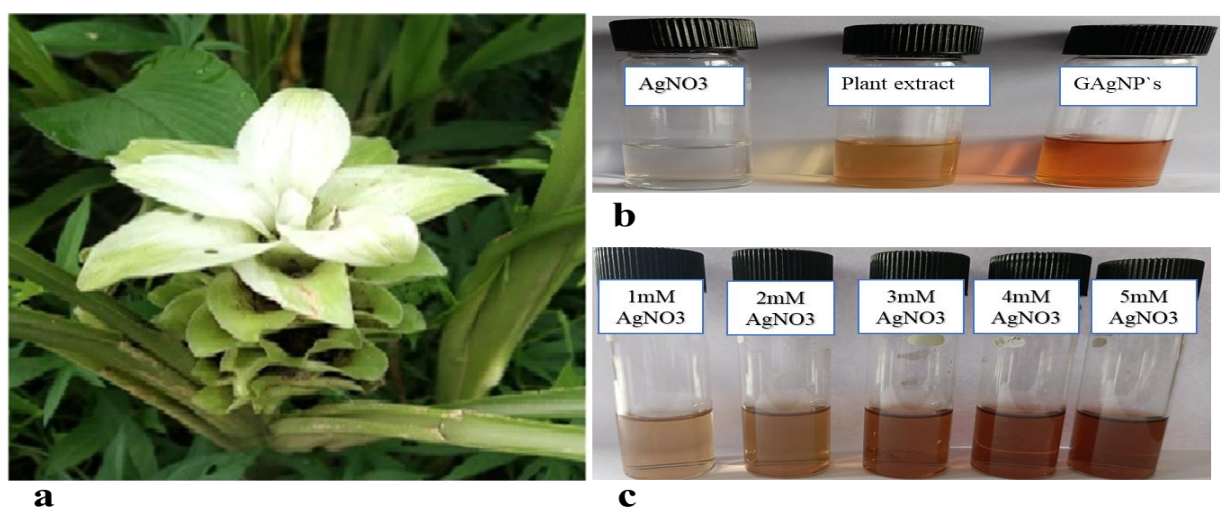

**Figure 1**. (**a**) *C. longa* flower. (**b**) Green synthesis of silver nanoparticle at 5 mM silver nitrate. (**c**) Green synthesis of silver nanoparticles with various concentrations of silver nitrate.

**Characterizations of GAgNP's.**

*UV-Visible Spectrophotometer Analysis.*

The UV-Visible spectroscopy is one of the most popular methods of determining properties and the extent of particle formation. In this experiment, UV-Visible spectrophotometry was also used to monitor GAgNP's synthesis at wavelengths ranging from 200 nm-800 nm. A single broad and strong UV-Visible spectrum at 400-450 nm was observed, as shown in **Figure 2 (a)** indicating the surface plasmon resonance (SPR) and polydispersity nature, as previously reported.[24] The SPR spectra of GAgNP's derived from the higher concentration of silver nitrate showed a sharper and strong absorption band at 400-450 nm, as shown in **Figure 2 (b)**. The SPR spectra of nanoparticles depend on their size, shape, interparticle interactions, and free electron density.[25] In this study, the Mie scattering theory based on UV-Visible spectra was employed to determine the mean particle diameter of synthesized GAgNP's.[25]

The equation used to estimate the particle size of GAgNP′s is as follows:

$$w = \frac{(\varepsilon_0 + 2n^2)cmu_F}{2N_c e^2 D}$$

where, $\omega$ is full width at half maxima of the peak which follows a Lorentz shape, and $\varepsilon_0$, $n$, $c$, $m$, $u_F$, $N_c$, $e$, and $D$ are the frequency-independent part of the complex form of the dielectric constant, refractive index of water, the velocity of light, the mass of the electron, electron velocity at the Fermi energy, number of electrons per unit volume, the electron charge and diameter of the particle respectively. The particle size of GAgNP's estimated from the above equation is ~4 nm. The inset **Figure 2 (b)** shows the log-normal size distribution of the synthesized silver nanoparticles. Anantharaman et al., also employed the Mie scattering theory to calculate the size of silver nanoparticle synthesis by *Andrographis paniculate*.[26]

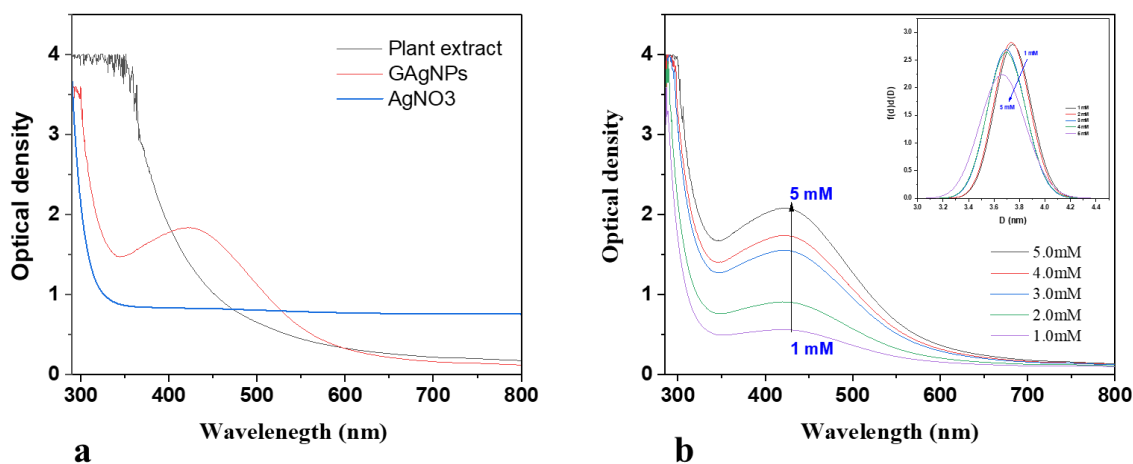

**Figure 2.** (**a**) UV–Visible absorption spectrum of silver nitrate, green synthesized silver nanoparticles, and crude extract. (**b**) UV-Visible spectra showing absorbance of green synthesized silver nanoparticles with different concentrations of silver nitrate and UV-Visible spectra, Inset: log-normal particle size distribution derived from UV spectra.

*XRD Analysis.*

XRD pattern analysis confirmed the crystalline nature of AgNP's synthesized using *C. longa*, see **Figure 3**. The XRD pattern, illustrated four characteristic peaks at 2θ values of 37.68°, 46.02°, 63.68°, and 76.64° corresponding to (111), (200), (220), and (311) planes respectively for silver nanoparticles, suggesting these nanoparticles were crystallized.[27] Based on the XRD pattern, Scherrer's equation was used to calculate the mean particle diameter of the synthesized GAgNP's:

$$D = (k\lambda/\beta \cos\theta)$$

A crystallite domain size is D in this equation, λ is the X-ray wavelength (1.5418 Å), β is the full width at half maximum (FWHM), k is known as the Scherer's constant (K=0.94), and θ is the diffraction angle.[28] According to the equation, GAgNP's has an average crystalline size of ~5 nm. There are few unidentified peaks (marked with stars) on the surface of the silver nanoparticles that indicate that the bio-organic phase has crystallized.[29] A similar result was reported in a study using *geranium* leaves.[30]

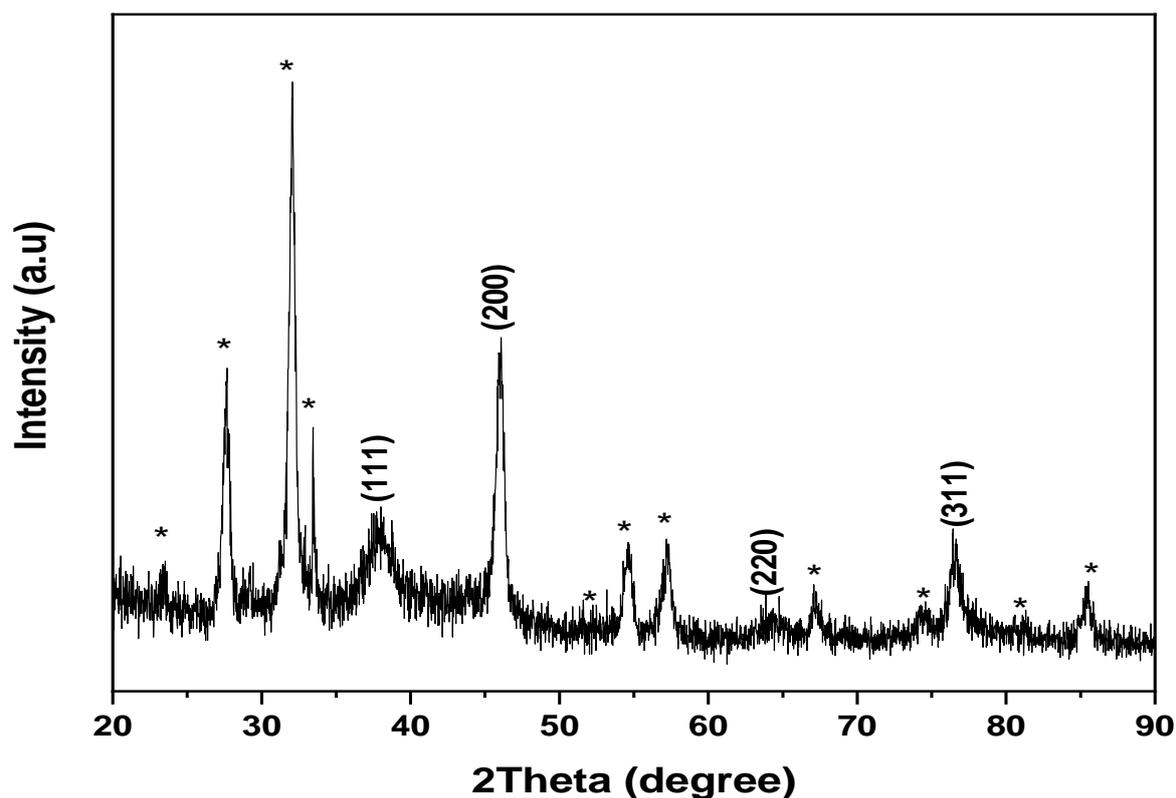

**Figure 3.** X-ray powder diffraction pattern of green synthesized silver nanoparticles.

*HR-TEM Analysis*

The size, shape, and morphology of nanoparticles were determined using HR-TEM. **Figure 4**, shows the HR-TEM analysis of green synthesized nanoparticles. The results demonstrate that most of the silver nanoparticles are well dispersed and spherical, and some are irregular in shape. The size distribution of the formed particles ranges from 3-14 nm with the majority of the particles being ~5 nm, which corroborates the X-ray diffraction data. The shape of green synthesized nanoparticles depends on the type of organic extract used.[31] Previously Nadagouda et al., showed that, in contrast to other types of plant extracts, turmeric powder in an aqueous solution could have good contact with gold and silver salts, which contributes to the formation of smaller nanoparticles.[32]

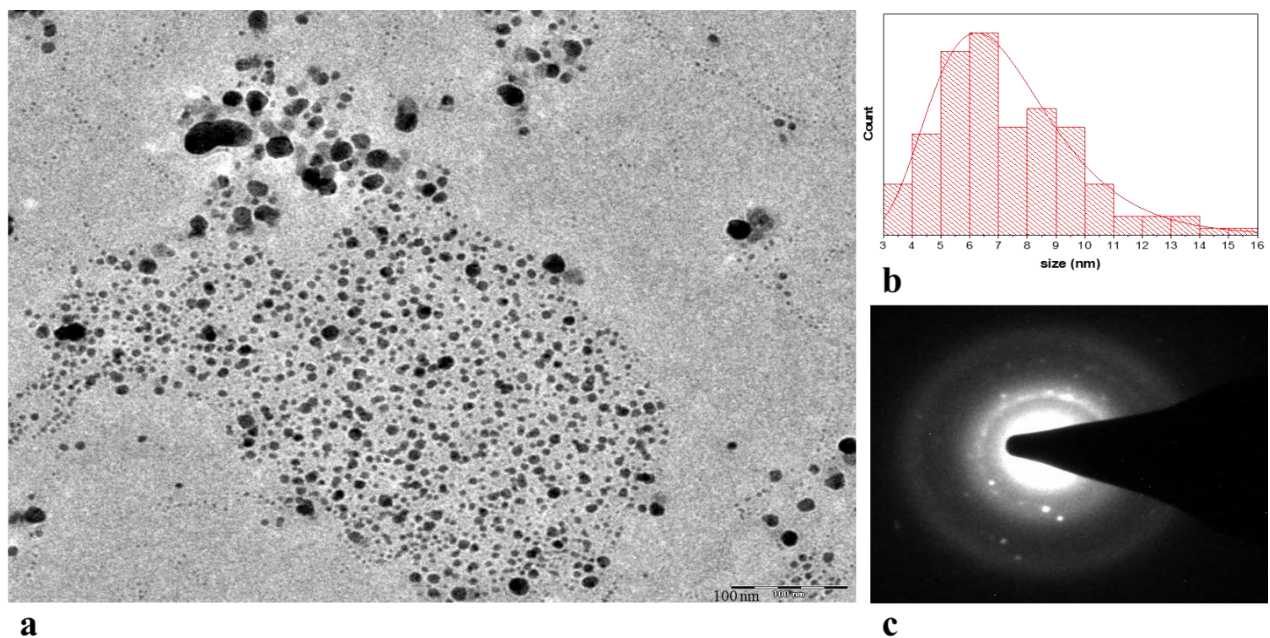

**Figure 4.** (**a**) HR-TEM image of GAgNP`s. (**b**) Particle size distribution histogram of GAgNP`s. (**c**) Power spectrum of nanoparticle sample subjected to HR-TEM.

The sum of all physical characterization of silver nanoparticles showed that the size of GAgNP's was found approximately to be around 5 nm. According to the study smaller size of silver nanoparticles enhanced the biological activity of nanoparticles.[33] This observation was supported by the antibacterial activity of chemically synthesized silver nanoparticles with 5 nm size which showed better antimicrobial activity against *E. coli*, *Fusobacterium nucleatum*, *Streptococcus mutants*, *Streptococcus sanguis*, *Streptococcus mitis*, *Aggregatibacter actinomycetemcomitans*, as compared to 15 nm and 25 nm silver nanoparticles.[34] Also, Agnihotri et al., observed and described that GAgNP's with less than 10 nm size possessed greater antibacterial efficacy; alongside, they concluded that GAgNP's of 5 nm size had the fastest antibacterial activity compared to other sizes.[35] All physical characterization confirmed that the synthesized silver nanoparticles were homogeneous with a size of ~5nm. Based on all reported studies, we further planned for the antimicrobial activity of the GAgNP's.

**Antibacterial Activity.**

*Disc Diffusion Assay.*
In response to the urgent need for the development of new antimicrobial reagents due to the emergence of multi-drug resistant (MDR) microorganisms which has resulted in an increase in

13the cost of medicine. In order to combat this problem, new bacteriocides development is mandated. Here we examined the bioactive properties of the aqueous floral extract of *C. longa* and the GAgNP's against six bacteria, including Gram-positives, Gram-negatives, and acid-fast bacteria such as *S. aureus*, *S. epidermidis*, *K. pneumoniae*, *E. coli*, *M. smegmatis*, and *M. phlei*. The results are presented in **Table 1**. and **Figure 5**. The biosynthesized GAgNP's demonstrated potential activity in comparison to the floral extracts. In comparison with the antibiotic controls, the synthesized GAgNP's showed significant activity (measured by the zone of inhibition) against all selected bacteria: 26 mm for *M. smegmatis*, 22 mm for *M. phlei* and *S. aureus*, 18 mm for *S. epidermidis* and *K. pneumoniae* and 13 mm against *E. coli*. Meanwhile, the 5 mm silver nitrate control and DMSO which was subjected as a control for the solvent system, exhibited no inhibitory effect against any of the selected bacteria.

**Table 1**. Result of primary antimicrobial screening with disc diffusion assay

| Bacteria | Zone of inhibition (in mm) | | |
|---|---|---|---|
| | GAgNP's | Plant extract | Antibiotics |
| *Staphylococcus aureus* | 22 | 0 | 20 |
| *Staphylococcus epidermidis* | 18 | 0 | 20 |
| *Escherichia coli* | 13 | 8 | 15 |
| *klebsiella pneumoniae* | 18 | 0 | 15 |
| *Mycobacterium smegmatis* | 26 | 6 | 30 |
| *Mycobacterium phlei* | 22 | 6 | 20 |

\* 0 did not show any growth inhibition.

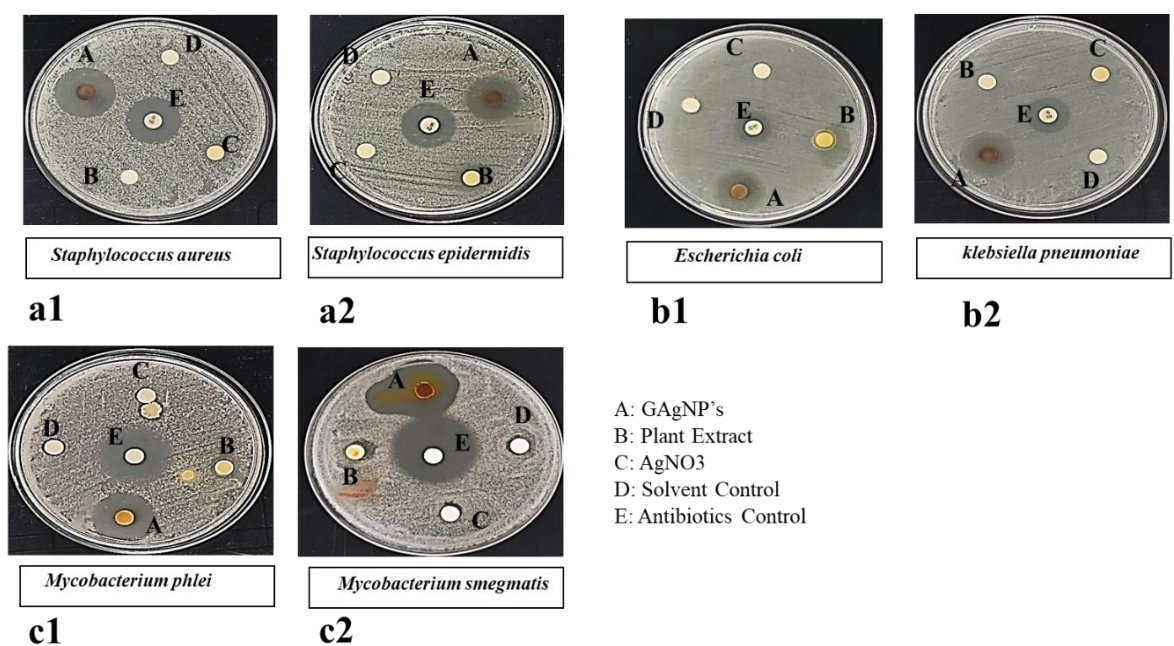

**Figure 5**. Anti-bacterial activity of *C. longa* flower extract and green synthesized silver nanoparticles by disc diffusion assay. Gram-positive bacteria (**a1** and **a2**), Gram-negative bacteria (**b1** and **b2**), and Acid-fast bacteria (**c1** and **c2**). GAgNP`s (**A**), Plant extract (**B**), AgNO$_3$ (**C**), Solvent control (**D**), Antibiotics control (**E**) (Streptomycin in **a1**, **a2**, **b1** and **b2** at 10 μg, Isoniazid in **c1** and **c2** at 10 μg).

*MIC Determination.*

The disk diffusion assay is considered as a preliminary evaluation to examine the antimicrobial activity of any designated agent. Moreover, GAgNP's antibacterial activity was further assessed based on their minimum inhibitory concentration (MIC) values. The MIC represents the lowest concentration of an antimicrobial that inhibits the visible growth of a microorganism after overnight incubation.[36] The MIC of synthesized GAgNP's were determined using the micro-broth dilution method, after the disc diffusion assay confirmed their antimicrobial activity as depicted in **Figure 6 (a)**. The MIC results correlate with the broader zone of inhibition, which directly corresponds to a smaller MIC, in other words, the larger diameter of the zone of inhibition in the disc diffusion assay would exhibit less MIC value.[37] Nonetheless, the MIC value of GAgNP's was comparatively lower than flower water extract against all selected Gram-positive, Gram-negative, and acid-fast bacteria, **Figure 6 (b)**. The observed antibacterial activity was attributed to GAgNP's. The MIC value for acid-fast bacteria *M. smegmatis* and *M. phlei* was 39.06 μg/mL. MIC values of 156.2 μg/mL and 78.12 μg/mL were obtained for Gram-negative bacteria *E. coli* and *K. pneumoniae*, respectively. For Gram-positive bacteria *S. aureus* and *S. epidermidis*, MIC values of 625 μg/mL and 78.12 μg/mL were obtained. The MIC values of the positive control drugs, streptomycin, and isoniazid, were found to be within the specified range for their corresponding pathogen. For instance, *S. aureus* and *E. coli* had MIC of 4 µg, *S. epidermidis*, *K. pneumoniae*, and *M. smegmatis* had MIC of 2 µg, and *M. phlei* had MIC of 0.5 µg.[38,39]

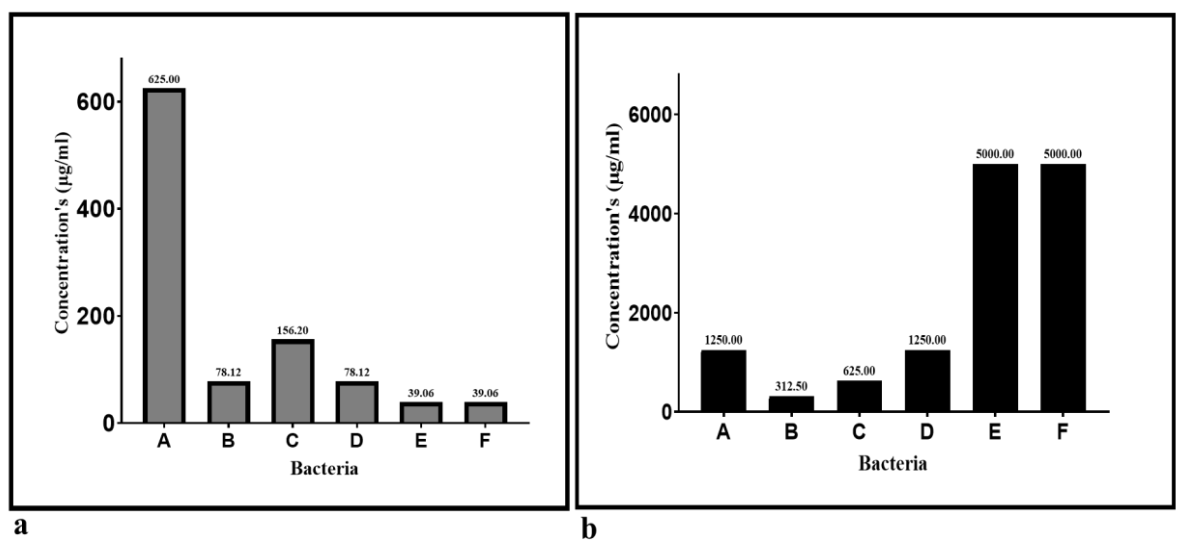

**Figure 6**. (**a**) Graph showing the MIC of GAgNP's against the specified bacteria. (**b**) Graph showing the MIC of *C. longa* flower extract against specified bacteria. *S. aureus* (**A**), *S. epidermidis* (**B**), *E. coli* (**C**), *K. pneumoniae* (**D**), *M. smegmatis* (**E**), *M. phlei* (**F**).

Silver nanoparticles are established as potent antimicrobial agents. Although significant progress has been achieved in the elucidation of the antimicrobial mechanism of silver nanoparticles, however, the exact mechanism of action is still unknown.[40] The positive charge confers electrostatic attraction between GAgNP's and the negatively charged cell membrane of the microorganisms, thereby facilitating GAgNP's attachment onto cell membranes.[41] The attachment of silver nanoparticles to the cell causes the degradation of the nanoparticle, to release $Ag^0$. Subsequently, $Ag^0$ possibly crosses the membrane through irregular pit formed on the cell wall due to nanoparticles, and also with the assistance of cation-selective porins.[42] After transport, the primary action of $Ag^0$ promotes the activation of ROS. This activation, in turn, leads to DNA damage, respiratory chain (ATP depletion) inactivation, membrane permeability disruption, cell cycle inhibition, and protein synthesis inhibition.[43] Based on the antimicrobial result analysis, the difference in the zone of inhibition and concertation of MIC against selected bacteria is possibly due to varying thicknesses and molecular compositions of the membrane structures of different bacteria. This variability may explain the differences in their sensitivities to GAgNP's.[44] According to Feng et al., the antimicrobial potential of silver nanoparticles may be reduced by thick cell wall and negative charge on the bacterial cell wall, which can cause the silver nanoparticles to become stuck on the cell wall.[45] In our study *Mycobacterium* was more sensitive to silver nanoparticles compared to Gram-positive and Gram-negative bacteria, possibly due to the differences in their cell wall permeability.[46] Results from these studies support the development of an efficient, biocompatible antibacterial molecule using metal nanoparticles.[47]

Based on several corroborations, GAgNP's have been shown to exhibit greater antimicrobial activity in comparison to flower extract. However, the biological activity of synthesized GAgNP's was of meagre standards (found to be sub-optimal in comparison to other reports) with respect to several reports. The hindrance in achieving better results could be due to certain factors such as plant metabolites sparsely detected during physical characterizations. Nevertheless, there's a possibility that using an appropriate agent to clean the GAgNP's for any unknown residues could improve the biological (antibacterial) activities, and further studies can yield better results.

- **CONCLUSION**

Green synthesized silver nanoparticles (GAgNP's) with reducing fundamentals of *C. longa* extract exhibited homogeneous with a size of ~5nm and significant antibacterial activity in the very sparse quantity of nanoparticles compared to the crude extract. Describe the bactericidal efficacy and can be considered a potential source of pharmaceutical conjugate and concoction for developing new antibacterial with the supporting physicochemical properties described can lead to a ray of new opportunities.

**Conflict of interest**

Author's hold no conflict of interest

- **ACKNOWLEDGEMENTS**

Authors acknowledge Charotar University of Science & Technology for providing suitable technical resources for conducting the current study; Alongside Dr. Dhara N. Patel, Head- Medical Laboratory Technology for providing clinical Isolates for the study. We acknowledge Prof. Rakesh Bhatnagar (Professor, School of Life Sciences, JNU, Delhi India) for gifting *M.*

*smegmatis* mc$^2$155. Author Kamal Kishor Rajak acknowledges the Department of Biotechnology (DBT) Government of India for providing the senior research fellowship.

- **REFERENCES**

(1) Mody, V. V; Siwale, R.; Singh, A.; Mody, H. R. Introduction to Metallic Nanoparticles. *J Pharm Bioallied Sci* **2010**, *2* (4), 282. https://doi.org/10.4103/0975-7406.72127.

(2) Iravani, S.; Korbekandi, H.; Mirmohammadi, S. V.; Zolfaghari, B. Synthesis of Silver Nanoparticles: Chemical, Physical and Biological Methods. *Res Pharm Sci* **2014**, *9* (6), 385.

(3) Seil, J. T.; Webster, T. J. Antimicrobial Applications of Nanotechnology: Methods and Literature. *Int J Nanomedicine* **2012**, *7*, 2767. https://doi.org/10.2147/IJN.S24805.

(4) Sharma, V. K.; Yngard, R. A.; Lin, Y. Silver Nanoparticles: Green Synthesis and Their Antimicrobial Activities. *Adv Colloid Interface Sci* **2009**, *145* (1–2), 83–96. https://doi.org/https://doi.org/10.1016/j.cis.2008.09.002.

(5) Thuesombat, P.; Hannongbua, S.; Akasit, S.; Chadchawan, S. Effect of Silver Nanoparticles on Rice (Oryza Sativa L. Cv. KDML 105) Seed Germination and Seedling Growth. *Ecotoxicol Environ Saf* **2014**, *104*, 302–309. https://doi.org/https://doi.org/10.1016/j.ecoenv.2014.03.022.

(6) Zhang, D.; Ma, X.; Gu, Y.; Huang, H.; Zhang, G. Green Synthesis of Metallic Nanoparticles and Their Potential Applications to Treat Cancer. *Front Chem* **2020**, *8*, 799. https://doi.org/https://doi.org/10.3389/fchem.2020.00799.

(7) Alqarni, L. S.; Alghamdi, M. D.; Alshahrani, A. A.; Nassar, A. M. Green Nanotechnology: Recent Research on Bioresource-Based Nanoparticle Synthesis and Applications. *J Chem* **2022**, *2022*. https://doi.org/https://doi.org/10.1155/2022/4030999.

(8) Guan, Z.; Ying, S.; Ofoegbu, P. C.; Clubb, P.; Rico, C.; He, F.; Hong, J. Green Synthesis of Nanoparticles: Current Developments and Limitations. *Environ Technol Innov* **2022**, 102336. https://doi.org/https://doi.org/10.1016/j.eti.2022.102336.

(9) Larue, C.; Castillo-Michel, H.; Sobanska, S.; Cécillon, L.; Bureau, S.; Barthès, V.; Ouerdane, L.; Carrière, M.; Sarret, G. Foliar Exposure of the Crop Lactuca Sativa to Silver Nanoparticles: Evidence for Internalization and Changes in Ag Speciation. *J Hazard Mater* **2014**, *264*, 98–106. https://doi.org/https://doi.org/10.1016/j.jhazmat.2013.10.053.

(10) Bar, H.; Bhui, D. K.; Sahoo, G. P.; Sarkar, P.; De, S. P.; Misra, A. Green Synthesis of Silver Nanoparticles Using Latex of Jatropha Curcas. *Colloids Surf A Physicochem Eng Asp* **2009**, *339* (1–3), 134–139. https://doi.org/https://doi.org/10.1016/j.colsurfa.2009.02.008.

(11) Das, R. K.; Brar, S. K. Plant Mediated Green Synthesis: Modified Approaches. *Nanoscale* **2013**, *5* (21), 10155–10162. https://doi.org/10.1039/C3NR02548A.

(12) Vadlapudi, V.; Kaladhar, D. Green Synthesis of Silver and Gold Nanoparticles. *Middle East J Sci Res* **2014**, *19* (6), 834–842. 10.5829/idosi.mejsr.2014.19.6.11585.

(13) Shipway, A. N.; Katz, E.; Willner, I. Nanoparticle Arrays on Surfaces for Electronic, Optical, and Sensor Applications. *ChemPhysChem* **2000**, *1* (1), 18–52.

https://doi.org/https://doi.org/10.1002/1439-7641(20000804)1:1<18::AID-CPHC18>3.0.CO;2-L.

(14) Sonvico, F.; Dubernet, C.; Colombo, P.; Couvreur, P. Metallic Colloid Nanotechnology, Applications in Diagnosis and Therapeutics. *Curr Pharm Des* **2005**, *11* (16), 2091–2105. https://doi.org/https://doi.org/10.2174/1381612054065738.

(15) Hebeish, A.; El-Rafie, M. H.; El-Sheikh, M. A.; Seleem, A. A.; El-Naggar, M. E. Antimicrobial Wound Dressing and Anti-Inflammatory Efficacy of Silver Nanoparticles. *Int J Biol Macromol* **2014**, *65*, 509–515. https://doi.org/https://doi.org/10.1016/j.ijbiomac.2014.01.071.

(16) Singh, A.; Kaur, K. Biological and Physical Applications of Silver Nanoparticles with Emerging Trends of Green Synthesis. *Engineered Nanomaterials-Health and Safety* **2019**. https://doi.org/10.5772/intechopen.88684.

(17) Reddy, A. V.; Suresh, J.; Yadav, H. K. S.; Singh, A. A Review on Curcuma Longa. *Res J Pharm Technol* **2012**, *5* (2), 158–165.

(18) Fuloria, S.; Mehta, J.; Chandel, A.; Sekar, M.; Rani, N. N. I. M.; Begum, M. Y.; Subramaniyan, V.; Chidambaram, K.; Thangavelu, L.; Nordin, R. A Comprehensive Review on the Therapeutic Potential of Curcuma Longa Linn. in Relation to Its Major Active Constituent Curcumin. *Front Pharmacol* **2022**, *13*. https://doi.org/10.3389/fphar.2022.820806.

(19) Verma, S. C.; Jain, C. L.; Rani, R.; Pant, P.; Singh, R.; Padhi, M. M.; Devalla, R. B. Simple and Rapid Method for Identification of Curcuma Longa Rhizomes by Physicochemical and HPTLC Fingerprint Analysis. *Chem Sci Trans* **2012**, *1* (3), 709–715. https://doi.org/10.7598/cst2012.203.

(20) Hemlata; Meena, P. R.; Singh, A. P.; Tejavath, K. K. Biosynthesis of Silver Nanoparticles Using Cucumis Prophetarum Aqueous Leaf Extract and Their Antibacterial and Antiproliferative Activity against Cancer Cell Lines. *ACS Omega* **2020**, *5* (10), 5520–5528. https://doi.org/10.1021/acsomega.0c00155.

(21) Patel, J. B.; Tenover, F. C.; Turnidge, J. D.; Jorgensen, J. H. Susceptibility Test Methods: Dilution and Disk Diffusion Methods. *Manual of clinical microbiology* **2011**, 1122–1143. https://doi.org/https://doi.org/10.1128/9781555816728.ch68.

(22) Wiegand, I.; Hilpert, K.; Hancock, R. E. W. Agar and Broth Dilution Methods to Determine the Minimal Inhibitory Concentration (MIC) of Antimicrobial Substances. *Nat Protoc* **2008**, *3* (2), 163–175. https://doi.org/https://doi.org/10.1038/nprot.2007.521.

(23) Daniel, M.-C.; Astruc, D. Gold Nanoparticles: Assembly, Supramolecular Chemistry, Quantum-Size-Related Properties, and Applications toward Biology, Catalysis, and Nanotechnology. *Chem Rev* **2004**, *104* (1), 293–346. https://doi.org/https://doi.org/10.1021/cr030698.

(24) Vinod, V. T. P.; Saravanan, P.; Sreedhar, B.; Devi, D. K.; Sashidhar, R. B. A Facile Synthesis and Characterization of Ag, Au and Pt Nanoparticles Using a Natural Hydrocolloid Gum Kondagogu (Cochlospermum Gossypium). *Colloids Surf B Biointerfaces* **2011**, *83* (2), 291–298. https://doi.org/https://doi.org/10.1016/j.colsurfb.2010.11.035.

(25) Desai, R.; Mankad, V.; Gupta, S. K.; Jha, P. K. Size Distribution of Silver Nanoparticles: UV-Visible Spectroscopic Assessment. *Nanoscience and nanotechnology letters* **2012**, *4* (1), 30–34. https://doi.org/https://doi.org/10.1166/nnl.2012.1278.

(26) Anantharaman, S.; Rego, R.; Muthakka, M.; Anties, T.; Krishna, H. Andrographis Paniculata-Mediated Synthesis of Silver Nanoparticles: Antimicrobial Properties and Computational

Studies. *SN Appl Sci* **2020**, *2* (9), 1–14. https://doi.org/https://doi.org/10.1007/s42452-020-03394-7.

(27) Khan, M.; Khan, M.; Adil, S. F.; Tahir, M. N.; Tremel, W.; Alkhathlan, H. Z.; Al-Warthan, A.; Siddiqui, M. R. H. Green Synthesis of Silver Nanoparticles Mediated by Pulicaria Glutinosa Extract. *Int J Nanomedicine* **2013**, *8*, 1507. https://doi.org/10.2147/IJN.S43309.

(28) Arokiyaraj, S.; Arasu, M. V.; Vincent, S.; Prakash, N. U.; Choi, S. H.; Oh, Y.-K.; Choi, K. C.; Kim, K. H. Rapid Green Synthesis of Silver Nanoparticles from Chrysanthemum Indicum L and Its Antibacterial and Cytotoxic Effects: An in Vitro Study. *Int J Nanomedicine* **2014**, *9*, 379. https://doi.org/10.2147/IJN.S53546.

(29) Vanaja, M.; Annadurai, G. Coleus Aromaticus Leaf Extract Mediated Synthesis of Silver Nanoparticles and Its Bactericidal Activity. *Appl Nanosci* **2013**, *3* (3), 217–223. https://doi.org/https://doi.org/10.1007/s13204-012-0121-9.

(30) Shankar, S. S.; Ahmad, A.; Sastry, M. Geranium Leaf Assisted Biosynthesis of Silver Nanoparticles. *Biotechnol Prog* **2003**, *19* (6), 1627–1631. https://doi.org/10.1021/bp034070w.

(31) Alsammarraie, F. K.; Wang, W.; Zhou, P.; Mustapha, A.; Lin, M. Green Synthesis of Silver Nanoparticles Using Turmeric Extracts and Investigation of Their Antibacterial Activities. *Colloids Surf B Biointerfaces* **2018**, *171*, 398–405. https://doi.org/https://doi.org/10.1016/j.colsurfb.2018.07.059.

(32) Nadagouda, M. N.; Iyanna, N.; Lalley, J.; Han, C.; Dionysiou, D. D.; Varma, R. S. Synthesis of Silver and Gold Nanoparticles Using Antioxidants from Blackberry, Blueberry, Pomegranate, and Turmeric Extracts. *ACS Sustain Chem Eng* **2014**, *2* (7), 1717–1723. https://doi.org/https://doi.org/10.1021/sc500237k.

(33) Szerencsés, B.; Igaz, N.; Tóbiás, Á.; Prucsi, Z.; Rónavári, A.; Bélteky, P.; Madarász, D.; Papp, C.; Makra, I.; Vágvölgyi, C. Size-Dependent Activity of Silver Nanoparticles on the Morphological Switch and Biofilm Formation of Opportunistic Pathogenic Yeasts. *BMC Microbiol* **2020**, *20* (1), 1–13. https://doi.org/https://doi.org/10.1186/s12866-020-01858-9.

(34) Lu, Z.; Rong, K.; Li, J.; Yang, H.; Chen, R. Size-Dependent Antibacterial Activities of Silver Nanoparticles against Oral Anaerobic Pathogenic Bacteria. *J Mater Sci Mater Med* **2013**, *24* (6), 1465–1471. https://doi.org/10.1007/s10856-013-4894-5.

(35) Agnihotri, S.; Mukherji, S.; Mukherji, S. Size-Controlled Silver Nanoparticles Synthesized over the Range 5–100 Nm Using the Same Protocol and Their Antibacterial Efficacy. *RSC Adv* **2014**, *4* (8), 3974–3983. https://doi.org/https://doi.org/10.1039/C3RA44507K.

(36) Andrews, J. M. Determination of Minimum Inhibitory Concentrations. *Journal of antimicrobial Chemotherapy* **2001**, *48* (suppl_1), 5–16. https://doi.org/https://doi.org/10.1093/jac/48.suppl_1.5.

(37) Benkova, M.; Soukup, O.; Marek, J. Antimicrobial Susceptibility Testing: Currently Used Methods and Devices and the near Future in Clinical Practice. *J Appl Microbiol* **2020**, *129* (4), 806–822. https://doi.org/https://doi.org/10.1111/jam.14704.

(38) Schatz, A.; Bugle, E.; Waksman, S. A. Streptomycin, a Substance Exhibiting Antibiotic Activity against Gram-Positive and Gram-Negative Bacteria. *Proceedings of the Society for Experimental Biology and Medicine* **1944**, *55* (1), 66–69. https://doi.org/10.3181/00379727-55-14461.

(39) Chaturvedi, V.; Dwivedi, N.; Tripathi, R. P.; Sinha, S. Evaluation of Mycobacterium Smegmatis as a Possible Surrogate Screen for Selecting Molecules Active against Multi-Drug Resistant Mycobacterium Tuberculosis. *J Gen Appl Microbiol* **2007**, *53* (6), 333–337. doi: 10.2323/jgam.53.333.

(40) Durán, N.; Durán, M.; De Jesus, M. B.; Seabra, A. B.; Fávaro, W. J.; Nakazato, G. Silver Nanoparticles: A New View on Mechanistic Aspects on Antimicrobial Activity. *Nanomedicine* **2016**, *12* (3), 789–799. https://doi.org/10.1016/j.nano.2015.11.016.

(41) Yu-sen, E. L.; Vidic, R. D.; Stout, J. E.; McCartney, C. A.; Victor, L. Y. Inactivation of Mycobacterium Avium by Copper and Silver Ions. *Water Res* **1998**, *32* (7), 1997–2000. https://doi.org/https://doi.org/10.1016/S0043-1354(97)00460-0.

(42) Slavin, Y. N.; Asnis, J.; Häfeli, U. O.; Bach, H. Metal Nanoparticles: Understanding the Mechanisms behind Antibacterial Activity. *J Nanobiotechnology* **2017**, *15* (1), 1–20. https://doi.org/https://doi.org/10.1186/s12951-017-0308-z.

(43) Dakal, T. C.; Kumar, A.; Majumdar, R. S.; Yadav, V. Mechanistic Basis of Antimicrobial Actions of Silver Nanoparticles. *Front Microbiol* **2016**, *7*, 1831. https://doi.org/https://doi.org/10.3389/fmicb.2016.01831.

(44) Kim, J. S.; Kuk, E.; Yu, K. N.; Kim, J.-H.; Park, S. J.; Lee, H. J.; Kim, S. H.; Park, Y. K.; Park, Y. H.; Hwang, C.-Y. Antimicrobial Effects of Silver Nanoparticles. *Nanomedicine* **2007**, *3* (1), 95–101. https://doi.org/https://doi.org/10.1016/j.nano.2006.12.001.

(45) Feng, Q. L.; Wu, J.; Chen, G. Q.; Cui, F. Z.; Kim, T. N.; Kim, J. O. A Mechanistic Study of the Antibacterial Effect of Silver Ions on Escherichia Coli and Staphylococcus Aureus. *J Biomed Mater Res* **2000**, *52* (4), 662–668. https://doi.org/https://doi.org/10.1002/1097-4636(20001215)52:4<662::AID-JBM10>3.0.CO;2-3.

(46) Bogatcheva, E.; Dubuisson, T.; Protopopova, M.; Einck, L.; Nacy, C. A.; Reddy, V. M. Chemical Modification of Capuramycins to Enhance Antibacterial Activity. *Journal of Antimicrobial Chemotherapy* **2011**, *66* (3), 578–587. https://doi.org/https://doi.org/10.1093/jac/dkq495.

(47) Gosheger, G.; Hardes, J.; Ahrens, H.; Streitburger, A.; Buerger, H.; Erren, M.; Gunsel, A.; Kemper, F. H.; Winkelmann, W.; Von Eiff, C. Silver-Coated Megaendoprostheses in a Rabbit Model—an Analysis of the Infection Rate and Toxicological Side Effects. *Biomaterials* **2004**, *25* (24), 5547–5556. https://doi.org/https://doi.org/10.1016/j.biomaterials.2004.01.008.